\documentclass{desyproc}

\begin{document}
\title{Electroweak radiative corrections and heavy top}

\author{{\slshape M. I. Vysotsky$^1$}\\[1ex]
$^1$ITEP, Moscow, Russia
 }

\contribID{smith\_joe}

\desyproc{DESY-PROC-2008-xx}
\acronym{HQP08} 
\doi  

\maketitle

\begin{abstract}

\end{abstract}

\section{Lecture 1: 5 steps}

Why are we going to speak about electroweak radiative corrections (ERC)
in the summer 2008?

\bigskip

Practical aspects: SM prediction of the value of $M_H$ from ERC:
$M_H =  (85 + 30 - 20)$ GeV.

\bigskip
 If
LHC  discovers a heavy Higgs boson, it will mean that new electroweak nonsinglet
particle(s) do exist.

\bigskip
Besides Higgs: if other particle(s) are discovered at LHC -- their contribution(s) to ERC  will be one of the first questions
you would like to analyze.

\bigskip
 So: ERC will be a hot topic at LHC.

\bigskip
Theoretical aspect: creation of the {\it renormalizable} theory of weak
interactions in the 60's is one of the greatest achievements of theoretical physics in the XX century.

\bigskip
 So: an educated person should know how to calculate {\it radiative
corrections} in GWS theory.

\bigskip
Why to discuss ERC at the School on ``Heavy Quark Physics''?

Usually (QED, QCD) heavy particle contributions to rad.corr. are
damped.

Muon magnetic moment \cite{BKH} :

 $\mu= e/(2 m_{\mu})[1 + \alpha/(2\pi)(Schwinger) + ... + (m_{\mu}/\Lambda_0)^2$
( Berestetskii, Krohin, Khlebnikov, 1956)].

\bigskip
By the way,
$\mu^{exp} = [1 + (1165920.8 \pm 0.6) 10^{-9}]e/(2 m_{\mu}) $,
and experimental uncertainty corresponds to $\Lambda_0 > 3$ TeV,
just LHC scale...

\bigskip

Nondecoupling.

In electroweak theory heavy particle contributions to radiative corrections are enhanced
because of Higgs mechanism of mass generation. Let us estimate leading term in meson-antimeson
transition amplitude \cite{MV} .



\begin{figure}[!htb]
\centering
\epsfig{file=box_1.epsi,width=6cm}
\end{figure}

t'Hooft-Landau gauge, $G_H=1/p^2$

\begin{eqnarray}
(m_t/\eta)^4 \int d^4 p p_{\alpha}
p_{\beta}/[(p^2-m_t^2)^2 p^4]=   \nonumber\\
\sim m_t^2/\eta^4 \sim G^2_F m_t^2  \nonumber
\end{eqnarray}

$M_Z = 91.188(2) GeV; \;\; M_W = 80.400(25) GeV;$
$\Gamma_Z, \;\; Br(Z \to l^+ l^-), \;\;  A_{FB}^{l,c,b}$

and other parameters of $Z$
are measured now with  the precision
better than 0.1 \% .

\bigskip
How large are radiative corrections?

\bigskip
$\delta \sim
\frac{g^2}{16\pi^2} = \frac{\alpha_W}{4\pi} = \frac{\alpha}{4\pi
\sin^2 \theta} \approx 0.2\%  \Longrightarrow $

\bigskip
one needs to take into account these corrections to deal with 
experimental data.

\bigskip

Brief reminder.

QED: \;\;\;\; $L(e_0, m_0)$

At one loop we get:

$$
e = e_0[1+ c_e \frac{\alpha}{\pi}\ln \frac{\Lambda^2}{m_e^2}] \; ,
\;\; m = m_0[1+c_m \frac{\alpha}{\pi} \ln \frac{\Lambda^2}{m_e^2}]
\; , \eqno{(1)}  $$

where $\Lambda$ is the ultraviolet cutoff.

$e$ and $m$ are measured with record precision and from (1) we get:

 $e_0 \equiv
e_0(\Lambda,m,e)$, $m_0 \equiv m_0(\Lambda,m,e)$.

The next step is the calculation of some amplitude; say Compton scattering,
$e\gamma \to e\gamma$:

 $$A = A_0(e_0, m_0) +
A_1(e_0, m_0, \Lambda)  =  A(e,m,s,t)$$

In this way we get a finite expression with one loop
radiative corrections taken into account.

 QED (quod erat demonstrandum).

\bigskip
$SU(2)_L\otimes U(1)$

The situation differs from QED.

\bigskip
Let us consider a gauge sector:

\bigskip
1. $L(g,g',\eta)$. These coupling constants and Higgs
expectation value are not measured directly
    and are known with rather poor precision;

2. $M_Z$ is known precisely, but it is not a parameter of the Lagrangian...

\bigskip
So, some modifications are needed.

Different approaches to study radiative corrections are possible.

\bigskip
1989 - start of SLC and LEP I;

\bigskip
LEPTOP,
1991 - 1995

 Victor Novikov, Lev Okun, Alexander
Rozanov, M.V. \cite{LT}

\bigskip

(ZFITTER (D.Yu. Bardin et al., Dubna - Zeuten) - was
widely used by LEP collaborations to deal with raw data;
other approaches can be found in the literature).

\bigskip
5 steps to heaven

1. The best measured observables:
$G_{\mu}$, $m_Z$,  $\alpha$

\bigskip
2. $G_{\mu} = G_{\mu}(g_0, \bar{g}_0,  \eta_0; \Lambda)$, \;\;\;
$m_Z =..., \;\;\; \alpha =...      $

\bigskip
3.  $g_0 = g_0(G_{\mu}, m_Z,
\alpha; \Lambda)$; \;\;\; $\bar{g}_0 = \bar{g}_0(G_{\mu},m_Z, \alpha; \Lambda)$;
$\eta_0 = \eta_0(G_{\mu},m_Z, \alpha; \Lambda)$

\bigskip
4. $m_W = m_W(g_0, \bar{g}_0, \eta_0; \Lambda)$,

\bigskip
5.   $m_W = m_W[g_0(G_{\mu},m_Z, \alpha; \Lambda),
\bar{g}_0(G_{\mu},m_Z, \alpha; \Lambda), \eta_0(G_{\mu},m_Z, \alpha;
\Lambda);
 \Lambda]$

Dependence on $\Lambda$ in the last expression cancels because
the theory is renormalizable.

Take other observables

($\Gamma_Z = ..., \;\;\;  A_{FB}= ...$,  \;\;\; ...)

and repeat items 4 and 5.

\bigskip
This is all what is needed to take into account electroweak 
radiative corrections at one loop.

\bigskip
QED.

\bigskip
A technical remark: ultraviolet cutoff $\Lambda$ breaks local gauge
invariance. To restore it in QED one subtracts photon mass, which
appears to be proportional to $e\Lambda$.

In QAD we wish to calculate
IVB masses.

\bigskip
The way out: dimensional regularization.

We will calculate integrals in  $D = 4 -2\varepsilon$ dimensional space-time,
where they converge and local gauge invariance is not spoiled.

\bigskip
So, in all formulas instead of $\Lambda$  poles $1/\varepsilon$ will occur.
In final formulas which express physical quantities
($M_W, \Gamma_Z,...$) through $G_{\mu}$, $m_Z$,  $\alpha$ these poles cancel.

\newpage
{step 2, $\alpha$}

\begin{figure}[!htb]
\centering
\epsfig{file=gamma_1.epsi,width=10cm}
\end{figure}

$$
\alpha = \frac{e_0^2}{4\pi}[1-\Pi'_{\gamma}(0) -2\frac{s}{c}\frac{\Pi_{\gamma
Z}(0)}{M_Z^2}] \;\; ,
\eqno{(2)}
$$

where  $Z$ interaction is given by

$\bar{g}(T_3-Qs^2) = -Q\frac{s}{c}e$, \;\; $\bar{g} = g/c = e/(cs)$

\bigskip
$s \equiv sin\theta$ - the sine of electroweak mixing angle.

\bigskip
$\alpha \to \bar{\alpha}$

The obtained equation for fine structure constant can be used to get
formulas for electroweak rad.corr.

However, since
 $\Pi'_{\gamma}(0) \sim \alpha \ln(\Lambda^2/m^2_{l,q})$,
where $m_l$ and  $m_q$ are the masses of charged leptons and quarks,
in final expressions u.v. cutoff $\Lambda$ will be substituted by
$M_Z$, and logarithmically enhanced rad. corr. will emerge.

Their physical sense is transparent: they correspond to $\alpha$
running from
 $q^2=0$ to the electroweak scale $q^2 =M_Z^2$.

\bigskip
It is very convenient to take this running into account from the very
beginning, separating it from proper weak rad. corr.

$$
\bar{\alpha} \equiv \alpha(M_Z) =
\frac{e_0^2}{4\pi}[1-\frac{\Pi_{\gamma}(M_Z^2)}{M_Z^2} - 2\frac{s}{c}
\frac{\Pi_{\gamma Z}(0)}{M_Z^2}] \;\;
\eqno{(3)}
$$
This equation will be used to determine the bare parameters of
electroweak Lagrangian  (remember that $e_0^2 = g_0^2(1-\frac{g_0^2}{\bar{g}_0^2})$).

\bigskip
From Eqs (2,3) one should find the numerical value of  $\bar{\alpha}$:

$$
\bar{\alpha} = \frac{\alpha}{1-\delta\alpha} \;, \;\; \delta\alpha =
\Pi'_{\gamma}(0) - \frac{\Pi_{\gamma}(M^2_Z)}{M_Z^2} \;\; ,
\eqno{(4)}
$$

and for electron loop one easily obtains:

$$
\delta\alpha = \frac{\alpha}{3\pi}\ln(\frac{M_Z^2}{m_e^2}) \;\; .
$$
Substituting it into Eq.(4) we obtain one of the most famous
equations in physics: zero charge formula of Landau, Abrikosov, Khalatnikov
\cite{LAK}.

\bigskip
Summing up leptonic and hadronic contributions we get:
$$
\alpha(M_Z) \equiv \bar{\alpha} =[128.95(5)]^{-1}
$$

instead of  $\alpha = [137.0359991(5)]^{-1}$.

\bigskip
step 2, $M_Z$

\begin{figure}[!htb]
\centering
\epsfig{file=zet_1.epsi,width=8cm}
\end{figure}

$$
G_{\mu\nu}^Z = \frac{-ig_{\mu\nu}}{k^2 -M_{Z_0}^2 +\Pi_Z(k^2)} + ... \;\; ,
$$

\bigskip
The pole position corresponds to the $Z$-boson mass:

$$
M_Z^2 = M_{Z_0}^2 -\Pi_Z(M_Z^2) \; , \;\;
M_{Z_0} = \frac{\bar{g}_0 \eta_0}{2} \;\; .
\eqno{(5)}
$$

\bigskip
step 2, $G_{\mu}$

$\mu \to e \nu_{\mu} \bar{\nu_e}$

\bigskip
It is convenient to divide rad. corr. into 2 parts:
dressing of $W$ - boson propagator, described by
 $\Pi_W(0)$, and vertexes and boxes, denoted by $D$:

$$
\frac{G_{\mu}}{\sqrt{2}} = \frac{g_0^2}{8m_{W_0}^2} [1+\frac{\Pi_W(0)}{M_W^2}
+D] = \frac{1}{2\eta_0^2}[1+\frac{\Pi_W(0)}{M_W^2}+D] \;\; .
\eqno{(6)}
$$

What about logarithmic running of the weak charge from
$q^2 \approx m_{\mu}^2$ to $q^2 \approx
M_W^2$?
 $\Pi_W(q^2)$
contains logarithmic term: $\Pi_W(q^2) \sim q^2 \ln
\frac{\Lambda^2}{max(q^2, m_e^2)}$.
However, due to nonzero mass of IVB running takes place
only above this mass.

\bigskip
So, there are two conditions for the charge to run logarithmically:

momentum transfer should be larger than the masses of
the particles in the loop
{\it and } larger than the mass of the corresponding vector boson.

Or the distances should be smaller...

\bigskip
That is why in the $Z$ - and $W$ - boson physics the big log occurs only in
the running of $\alpha$.

\bigskip
step 3

$$ G_{\mu} =
\frac{1}{\sqrt{2}\eta_0^2}[1+\frac{\Pi_W(0)}{M_W^2}+D] \;\; ,
$$
$$
M_Z^2
= \frac{1}{4}\bar{g}_0^2\eta_0^2 -\Pi_Z(M_Z^2) \;\; ,
$$ $$
4\pi\bar{\alpha} =g_0^2(1-\frac{g_0^2}{\bar{g}_0^2})
[1-\frac{\Pi_{\gamma}(M_Z^2)}{M_Z^2} -2\frac{s}{c}
\frac{\Pi_{\gamma Z}(0)}{M_Z^2}]  $$.

\bigskip
For bare parameters we get:
$$
\eta_0^2 =\frac{1}{\sqrt{2}G_{\mu}}[1+\frac{\Pi_W(0)}{M_W^2}+D] \;\; ,
$$
$$
\bar{g}_0^2 =4\sqrt{2}G_{\mu}M_Z^2[1+\frac{\Pi_Z(M_Z^2)}{M_Z^2}
-\frac{\Pi_W(0)}{M_W^2} -D] \;\; ,  $$

and it is convenient to rewrite the equation for $g_0$ in the following way:

$$ \frac{g_0^2}{\bar{g}_0^2}(1-\frac{g_0^2}{\bar{g}_0^2}) =
\frac{\pi\bar{\alpha}}{\sqrt{2}G_{\mu}M_Z^2}
(1+\frac{\Pi_W(0)}{M_W^2} - \frac{\Pi_Z(M_Z^2)}{M_Z^2} +
\frac{\Pi_{\gamma}(M_Z^2)}{M_Z^2} + $$ $$ + 2\frac{s}{c}
\frac{\Pi_{\gamma Z}(0)}{M_Z^2} +D) \;\; . $$

\bigskip
$sin\theta$

Arithmetic was enough to solve the first 2 equations;
for the third one trigonometry is needed.

Let us define an electroweak mixing angle:

$$
\sin^2\theta
\cos^2\theta = \frac{\pi\bar{\alpha}}{\sqrt{2}G_{\mu}M_Z^2} \; ,
\;\; \sin^2\theta = 0.2310(1)\; $$

and solve the third equation:
$$
\frac{g_0}{\bar{g}_0}
=c[1+\frac{s^2}{2(c^2 -s^2)} (\frac{\Pi_Z(M_Z^2)}{M_Z^2} -
\frac{\Pi_W(0)}{M_W^2} - \frac{\Pi_{\gamma}(M_Z^2)}{M_Z^2}$$ $$ -2
\frac{s}{c} \frac{\Pi_{\gamma Z}(0)}{M_Z^2} -D)]  \;\; .
$$

\bigskip
step 4; custodial symmetry

Let us start from the $W$ - boson mass:

$$
M_W^2 = M_{W_0}^2 -\Pi_W(M_W^2) \; , \;\;
M_{W_0} = \frac{g_0 \eta_0}{2} \;\; .
$$

At step 2 an analogous equation was written for $M_Z$;
using it we get:

$$
\frac{M_W}{M_Z} = \frac{g_0}{\bar{g}_0} [1+\frac{\Pi_Z(M_Z^2)}{2M_Z^2} -
\frac{\Pi_W(M_W^2)}{2M_W^2}] \;\; .
$$
If
$U(1)$ charge $g'_0$ were zero,
then $g_0 = \bar{g}_0$, and at tree level $M_W = M_Z$,
which is a good approximation to the real life: $80 GeV \approx 90 GeV$.

\bigskip
$t$: {\it anti}custodial symmetry

What about loops? If $m_{up} = m_{down}$, then
$\Pi_Z(M_Z^2) = \Pi_W(M_W^2)$ and IVB stay degenerate.

In the real life top quark is extremely heavy, and the contribution
of the $(t,b)$ doublet to the difference of $W$- and $Z$- boson
masses is enhanced as
$m_t^2/M_Z^2
\approx 4$.

\bigskip
step 5, $ M_W $

$$
\frac{M_W}{M_Z} = c + \frac{c^3}{2(c^2 -s^2)}(\frac{\Pi_Z(M_Z^2)}{M_Z^2} -
\frac{\Pi_W(M_W^2)}{M_W^2}) +
$$
$$
+\frac{cs^2}{2(c^2 -s^2)}(
\frac{\Pi_W(M_W^2)}{M_W^2}
-\frac{\Pi_W(0)}{M_W^2} -\frac{\Pi_{\gamma}(M_Z^2)}{M_Z^2} -
$$
$$
-2\frac{s}{c}
\frac{\Pi_{\gamma Z}(0)}{M_Z^2} - D) \;\; , $$

UV divergences cancel in the last expression:

{\it the formula for finite one loop ew rad. corr. to the ratio of IVB masses is obtained!!!}

\bigskip
$Z \to l^+l^-$
Reminding that
$Z$-boson coupling constant is $\bar{g}_0$
and corresponding generator equals

$T_3 - s_0^2 Q$

we get for the decay amplitude at the tree level:

$$
A_0 = \frac{\bar{g}_0}{2} \bar{l}[-\frac{1}{2}\gamma_{\alpha}\gamma_5 -
(\frac{1}{2} -2s_0^2)\gamma_{\alpha}]l \; Z_{\alpha} \;\; .
$$
Taking into account the expressions for $\bar{g}_0$
and $g_0/\bar{g}_0$ as well as the loop corrections to
the tree diagram  we
straightforwardly obtain
the expression for the decay amplitude free from
the ultraviolet
divergences.

\begin{figure}[!htb]
\centering
\epsfig{file=zee_1.epsi,width=8cm}
\end{figure}

$$
A = \sqrt{\sqrt{2}G_{\mu}M_Z^2}[1+\frac{\Pi_Z(M_Z^2)}{2M_Z^2} -
\frac{\Pi_W(0)}{2M_W^2} - \frac{1}{2}D -
$$
$$
-\frac{1}{2}\Pi'_Z(M_Z^2)] \times
[ (-\frac{1}{2}+F_A) \bar{l}\gamma_{\alpha}\gamma_5 l +
$$
$$
+ \left (2s^2
-\frac{1}{2} +F_V + 2cs \frac{\Pi_{Z\gamma}(M_Z^2)}{M_Z^2} + \frac{2s^2
c^2}{c^2 -s^2} \times
\right.
$$
$$
\left. \times (\frac{\Pi_W(0)}{M_W^2} - \frac{\Pi_Z(M_Z^2)}{M_Z^2} +
\frac{\Pi_{\gamma}(M_Z^2)}{M_Z^2} +2\frac{s}{c} \frac{\Pi_{\gamma
Z}(0)}{M_Z^2} +D) \right
)$$
$$
\bar l \gamma_{\alpha}l]\; ,
$$
where functions $F_A$ and $F_V$ take into account corrections to $Zll$ vertex.

\bigskip
step 5, $g_A$ and $g_V$

Let us rewrite the expression for the decay amplitude:

$$
A = \sqrt{\sqrt{2}G_{\mu}M_Z^2} \bar{l}[g_A \gamma_{\alpha}\gamma_5 +g_V
\gamma_{\alpha}] l \; Z_{\alpha} \;\; .
$$

The UV finite expressions for axial and vector coupling constants
are given by a long formula above.

\section{Lecture 2:
 top and Higgs}

Formulas for electroweak radiative corrections
obtained in the first lecture are characterized  by strong
dependence on the top quark mass which helped to find the
top at TEVATRON in 1994.

There are two places in ew corrections to
IVB parameters where top quark contributions are enhanced:

1. the polarization operators (of nonconserved currents);

2. $Z \to t \bar t  \to b \bar b$ decay amplitude.

\bigskip
Why does current nonconservation matter?

 $\Pi_{\gamma}(q^2)$, $\Pi_{\gamma Z}(q^2) \sim [g_{\mu\nu}q^2
-q_{\mu}q_{\nu}](a + b q^2/m^2_t + ...)$,

while  $\Pi_W \sim g_{\mu\nu}(m_t^2 +...)$.

\bigskip
$m_t^2$ term from $\Pi$'s

In the limit $m_t^2 \gg m_W^2, m_Z^2$ we have:

 $\Pi(m_V^2) =\Pi(0)$, that is why we get the
following relations:

$$
\frac{M_W}{M_Z} = c+\frac{c^3}{2(c^2 -s^2)} (\frac{\Pi_Z(0)}{M_Z^2} -
\frac{\Pi_W(0)}{M_W^2}) \;\; ,
$$
$$
g_A = -\frac{1}{2} -
\frac{1}{4}(\frac{\Pi_Z(0))}{M_Z^2}-\frac{\Pi_W(0)}{M_W^2}) \;\; ,
$$
$$
g_V/g_A = 1-4s^2 + \frac{4c^2 s^2}{c^2 -s^2}
(\frac{\Pi_Z(0)}{M_Z^2}-\frac{\Pi_W(0)}{M_W^2}) \;\; .
$$

In order to honestly calculate the $m_t$ dependence of the
physical observables one should calculate the top quark
contributions to polarization operators:
$$
-i \Pi_{\mu\nu}^\psi(q^2) =-
$$
$$
-\int \frac{d^D k}{(2\pi)^D \mu^{D-4}} \frac{Sp
\gamma_{\mu}(\gamma_5)(\hat{k}+m_1)\gamma_\nu(\gamma_5)(\hat{k}+\hat{q}+m_2)}
{(k^2 -m_1^2)((k+q)^2 -m_2^2)} \;\; ,
$$
where we use dimensional regularization of the quadratically
divergent expression: $D = 4 - 2\varepsilon$ and the factor $\mu$
takes care of the canonical dimension of the integral.

\begin{figure}[!htb]
\centering
\epsfig{file=zw_1.epsi,width=8cm}
\end{figure}

``Back of the envelope'' calculation of $m_t^2$ term:

\bigskip
$$
\frac{\Pi_Z^\psi(0)}{M_Z^2} - \frac{\Pi_W^\psi(0)}{M_W^2} =
\frac{3 \bar\alpha}{8\pi c^2 s^2 M^2_Z}\int\frac{d p^2}
{(p^2 + m^2_t)^2} \times
$$

$$
\times \left[ p^4/2 + (p^2 + m_t^2)^2/2 - p^2(p^2 + m_t^2)\right] =
$$
$$
= \frac{3\bar{\alpha}}{16\pi c^2 s^2}
(\frac{m_t}{M_Z})^2 \;\; .
$$

\bigskip
Specific for $Z \to bb$ decay $m^2_t$ term comes from:

\begin{figure}[!htb]
\centering
\epsfig{file=zbb.epsi,width=9cm}
\end{figure}


\bigskip
Htb vertex is proportional to $m_t$, that is why one-loop
diagrams produce correction to $Zbb$ coupling enhanced as
$(m_t/M_Z)^2$.

To calculate this correction we can neglect $Z$-boson momentum.

$Z$-boson coupling is proportional to $T_3 - Q s^2$. The part
proportional to $b$-quark electric charge $Q$ induces vector
coupling which is not renormalized by Higgs loop (CVC) - so, at
zero momentum transfer ($q^Z_{\mu} = 0$) the sum of one loop
diagrams is zero.

What remains is $T_3$ which induces the coupling with $b_L$
and $t_L$.

And again the vector part is not renormalized, so only
axial current remains.

Since $Z$-boson has only vector coupling with Higgs, we should not
calculate corresponding vertex diagram. And  only the
diagram with $Ztt$ coupling should be taken into account.

Calculating a vertex diagram with UV  cutoff $\Lambda$  we get:
$$
\bar g/4/(16\pi^2)(\frac{m_t}{\eta/\sqrt{2}})^2\times
$$
$$
\times[-1/2 ln(\Lambda^2/m_t^2) + 3/2]
\bar b \gamma_{\alpha}\frac{(1 + \gamma_5)}{2}b\;Z_{\alpha}\;,
$$
while ($tH$) insertions into external legs give:
$$
\bar g/4/(16\pi^2)(\frac{m_t}{\eta/\sqrt{2}})^2\times
$$
$$
\times[1/2 ln(\Lambda^2/m_t^2) + 1/2]
\bar b \gamma_{\alpha}\frac{(1 + \gamma_5)}{2}b\;Z_{\alpha}\;.
$$

The sum of the two last expressions produces correction to
$g^b_L$:
$$
- \bar g /2[1 - (\frac{m_t}{\eta/\sqrt{2}})^2/(16\pi^2)]\bar b \gamma_{\alpha}\frac{(1 + \gamma_5)}{2}b\;Z_{\alpha}=
$$
$$
= - \bar g /2[1 - \frac{\alpha}{8\pi c^2 s^2}(\frac{m_t}{M_Z})^2] b \gamma_{\alpha}\frac{(1 + \gamma_5)}{2}b\;Z_{\alpha}\;,
$$

which reduces $\Gamma_Z(bb)$.

\bigskip
$M_H$

Electroweak rad. corr. depend on $M_H$. This is the reason
why from precision measurement of $Z$- and $W$-boson
 parameters
the value of $M_H$ is extracted.

\bigskip
Which diagrams matter? The radiation of Higgs from
the fermion line is
proportional to $m_f/\eta$, and since $\eta = 1/(\sqrt{\sqrt{2}
G_{\mu}}) = 246 GeV$ even in the case of $b$-quark it should be neglected.

What remain are the vector boson polarization operators (just as in
the case of top, if we forget for the moment $Z\to bb$ decay).

It is convenient to perform calculations in unitary gauge
where the nonphysical degrees of freedom ($H^{\pm}, Im H^0$)
are absent.

\begin{figure}[!htb]
\centering
\epsfig{file=h_1.epsi,width=8cm}
\end{figure}

The following substitution allows to extract Higgs interactions
with W-boson from W mass term:


$$
(\frac{g\eta}{2})^2|W|^2  \to
(\frac{g(\eta +H^0)}{2})^2|W|^2  =
$$
$$
=(\frac{g\eta}{2})^2|W|^2  +\frac{1}{2}
g^2 \eta H^0|W|^2 +\frac{g^2}{4} H^{0^2}|W|^2  =
$$
$$
= M_W^2|W|^2 +g M_W H^0|W|^2 +\frac{1}{4}g^2 H^{0^2}|W|^2
 \;\; .
$$

Analogously from the $Z$-boson mass term in Lagrangian we can
obtain $HZZ$ and $HHZZ$ coupling constants.

A less trivial example is the extraction of $H\gamma\gamma$
vertex from the following one loop effective Lagrangian:

$$L_{eff}(q^2 \ll M^2) = \frac{1}{4}F_{\mu\nu}^2\times
$$
$$
\times \sum\frac{b_ie^2}{16\pi^2}log(\frac{\Lambda^2}{M^2_i})\;\;,
$$
where $b = -4/3$ for a charged lepton, $-4 Q^2$ for a (colored) quark,
$7$ for $W$-boson are the coefficients of Gell-Mann -- Low
function. Substituting $M = f(g)\eta(1 + H/\eta)$ and expanding
the
logarithm we obtain the amplitude of $H\to \gamma\gamma$
decay. The most remarkable in the last formula is the sign
of the $W$ loop contribution, opposite to that of the lepton
and quark loop, and number 7 as well.

\bigskip
Asymptotic freedom in the USSR, 1965.

Let me start from number 7 obtained by V.S.Vanyashin
and M.V.Terentiev in their 1965 ZhETPh paper \cite{VT} .
At present the easiest way to derive it is the following:

$
7=11/3 C_V -1/6 -1/6, C_V = 2 $ for $SU(2) \;\; ,
$

where $11/3 C_V$ is the contribution of the massless vectors
in adjoint representation to the $\beta$-function. One factor $1/6$ comes from the Higgs doublet contribution to
the same SU(2) $\beta$-function,
 while another
$1/6$ is the Higgs doublet contribution to the running
of the coupling constant $g'$,
$$
1/e = 1/g + 1/g' \;\;.
$$

Concerning the sign Vanyashin and Terentiev stressed that it is
{\it opposite} to that which {\it always} occurs in QFT.

\bigskip
A bit of history: I heard from  Terentiev that
when he was giving a talk at ITEP  seminar
on this paper Pomeranchuk said that evidently the theory was
not selfconsistent (he relied upon Landau - Pomeranchuk
``zero-charge'' theorem). At the end of the paper this ``wrong sign'' behavior
of $\beta$-function was attributed  to
nonrenormalizability of the electrodynamic of massive charged vector bosons.

However in the Abstract the anomalous
character of the electric charge renormalization was
emphasized.

\bigskip
M.V. Terentiev worked at ITEP, V.S. Vanyashin worked and
is still
working at Dnepropetrovsk Physico-Technical Institute.

\bigskip
It is remarkable that if Higgs boson mass is
around 120 GeV (which is quite probable: SM fit, see below)
than $H \to \gamma\gamma$ decay will
play the important role in Higgs discovery and  factor ``7''
will become known to everybody in hep community 45 years
after its first appearance.

\bigskip
Back to Higgs in radiative corrections.

Calculation of $W$- and $Z$-boson polarization
operators provides us with explicit dependence of physical
observables on $M_H$. In the  limit
$M_H \gg M_Z$ we get:
$$
\Pi^H_W \sim M_H^2 \ln(M_H^2) + M_H^2 + (M^2_W + q^2)\ln(M_H^2)\;.
$$
In the differences of polarization operators on which
physical quantities depend:
$$
\frac{\Pi_W(M_W^2)}{M_W^2} - \frac{\Pi_W(0)}{M_W^2} \; ,
\;\;
\frac{\Pi_Z(M_Z^2)}{M_Z^2} - \frac{\Pi_W(M_W^2)}{M_W^2} \;
\;\;
\;\; \mbox{\rm and}
$$
$$
\Pi'_Z(M_Z^2)
$$

 the first two terms cancel
and we are left with
the logarithmic dependence on
Higgs mass\;.

\bigskip
Heavy top and Higgs asymptotics:

$$
\frac{M_W}{M_Z} = c+\frac{3\bar\alpha c}{32\pi s^2(c^2 -s^2)} \left[(\frac{m_t}{M_Z})^2 -
\frac{11}{9} s^2 \ln(\frac{M_H}{M_Z})^2 \right]  ,
$$
$$
g_A = -\frac{1}{2}-\frac{3\bar\alpha}{64\pi c^2 s^2}
\left[(\frac{m_t}{M_Z})^2 -s^2 \ln(\frac{M_H}{M_Z})^2\right]  ,
$$
$$
\frac{g_V}{g_A} = 1-4s^2 +\frac{3\bar\alpha}{4\pi(c^2 -s^2)}\left[(\frac{m_t}{M_Z})^2
-(s^2 +\frac{1}{9})\times \right.
$$
$$
\times \left. \ln(\frac{M_H}{M_Z})^2\right]  .
$$
Since the coefficients multiplied by log are almost equal, without the
knowledge of the value of $m_t$ one could not determine the value
of $M_H$.

\section{Lecture 3: SM fits; NP contributions}

After top discovery at Tevatron in 1994 the electroweak precision data provide information on Higgs mass.

The dependence on $M_H$ is provided by $\Pi$'s; the ``constants''
are also very important. The expressions in square
brackets at a previous slide are substituted by three
functions:
$$
V_m(t,h)\;, V_A(t,h)\;, V_R(t,h)\;;
$$
$$
 t \equiv (m_t/M_Z)^2,
h \equiv (M_H/M_Z)^2,
$$
which take into account all the existing loop calculations
($\alpha_W, \alpha_s\alpha_W$,..., for details see Novikov, Okun,
Rozanov, Vysotsky \cite{LT} .

\bigskip
Yellow Report

After the first years of LEPI operation it has become
clear that the
experimental data on $Z$ parameters will have very high
accuracy. That is why 4  codes which existed in literature
have been compared with
the aims to check numerical consistency of
the different approaches to radiative corrections calculation
and to determine
theoretical uncertainties.

The results are published in the
CERN Yellow Report 95-03, Editors D.Bardin, W. Hollik,
G.Passarino \cite{BHP} .

\bigskip
From history to our days.

LEPTOP fit of the precision observables
 (A.Rozanov, Summer 2008).

\begin{center}


\begin{tabular}{|l|l|l|r|}
\hline Observable & Exper. data  & LEPTOP fit & Pull \\ \hline
$\Gamma_Z$, GeV &    2.4952(23) &  2.4963(15)  & -0.5
\\ $\sigma_h$, nb &   41.540(37)& 41.476(14)  & 1.8    \\ $R_l$ &
20.771(25)& 20.743(18)  & 1.1   \\ $A_{\rm FB}^l$ & 0.0171(10)  &
0.0164(2)  & 0.8   \\ $A_{\tau}$ & 0.1439(43) &  0.1480(11)  &
-0.9   \\ $R_b$ &    0.2163(7) &   0.2158(1)  & 0.7   \\ $R_c$ &
0.172(3)&   0.1722(1)  & -0.0   \\ $A_{\rm FB}^b$  & 0.0992(16)&
0.1037(7)  & -2.8   \\ $A_{\rm FB}^c$  &    0.0707(35)& 0.0741(6)
& -1.0   \\ $s_l^2$ ($Q_{\rm FB}$)   &    0.2324(12)  & 0.2314(1)
& 0.8   \\
\hline
\end{tabular}
\end{center}

\begin{center}

\bigskip

\begin{tabular}{|l|l|l|r|}
\hline Observable & Exper. data  & LEPTOP fit & Pull \\
\hline $A_{\rm LR}$ &  0.1513(21)&
{0.1479(11)}  & 1.6   \\ $A_b$ &  0.923(20) & 0.9349(1)  & -0.6
\\ $A_c$ &  0.670(27)&   0.6682(5)  & 0.1   \\ \hline $m_W$, GeV &
80.398(25) & 80.377(17)  & 0.9   \\ \hline $m_t$, GeV    &
172.6(1.4) &   { 172.7(1.4)} &-0.1\\ $M_{\rm H}$, GeV  &   &  {
$84^{+32}_{-24} $} &  \\ $\hat{\alpha}_s$ & &  0.1184(27) &  \\
$1/\bar{\alpha}$ & 128.954(48) & 128.940(46) & 0.3  \\ {\small
$\chi^2/n_{\rm d.o.f}$} & & 18.1/12 & \\ \hline
\end{tabular}
\end{center}

With 10 MeV experimental accuracy of $M_W$ the accuracy in $M_H$ will be (+20
-15) GeV, while at the moment $M_H <140(150)$GeV at 95\% C.L.,
$M_H <185(200)$GeV at 99.5\% C.L.

(numbers in brackets take into account theoretical uncertainty).

This is the end of the Standard Model story.

\bigskip
The main results for other domains of particle physics:

\bigskip
1. QCD: power corrections for $Z$ width into hadrons
are definetly negligible; the obtained value of $\hat\alpha_s$
appears to be considerably larger than (some) QCD people
believed; in particular $J/\psi$ is outside of the perturbative
QCD domain;

2. GUT: the precise determination of $sin\theta$
excludes simplest SU(5) unification
theory without low energy SUSY.

\bigskip
New Physics.

What if LHC after a couple of months of operation will
announce the discovery of 300 GeV (or even heavier) Higgs?

It will definitely mean that  beyond Standard Model there are
other electroweak nonsinglet particles which contribute to the
functions $V_i$ and shift the value of $M_H$ in the minimum
of $\chi^2$.

Before discussing New Physics contribution to
radiative corrections let me present
two popular sets of parameters widely used in literature.

\bigskip
$\varepsilon_1, \varepsilon_2, \varepsilon_3$

A set of three parameters $\varepsilon_i$ has been suggested
by Altarelli, Barbieri and Jadach for the most general
phenomenological analysis of New Physics \cite{ABJ}. These parameters
are in one-to-one correspondence with our parameters $V_i$:

$$ \varepsilon_1 \sim \alpha_W V_A$$
$$ \varepsilon_2 \sim \alpha_W [(V_A - V_m) - 2 s^2
(V_A - V_R)]$$
 $$\varepsilon_3 \sim \alpha_W (V_A - V_R)$$

Since $ \varepsilon_2$ and $ \varepsilon_3$
do not contain the leading $\sim m^2_t$ term, their values
were useful in search for New Physics before the mass of
top quark was measured directly.

\bigskip
$S, T, U $

These letters, popular in particle physics, were used by
Peskin and Takeuchi for the parametrization of the so-called
oblique corrections due to New Physics
contribution to electroweak observables \cite{PT}.
Schematically:

$$\delta\varepsilon_1 = \alpha T\;,\;\;
\delta\varepsilon_2 \sim \alpha U\;,\;\;
\delta\varepsilon_3 = \alpha S\;,$$
where $\delta$ means that only NP contributions should be
taken into account.

Literally, Peskin and Takeuchi made one more step. Discussing NP
with a scale much larger than $M_Z$ they expanded the polarization
operators at $q^2 = 0$, taking into account only the first two
terms, $\Pi(0)$ and $\Pi'(0)$, which is the correct approximation
as far as higher derivatives are suppressed as
$[(M_Z^2)/(M_{NP}^2)]^n$. One can find in the literature (PDG) the
allowed domains of $S,U$ and $T$ for a given value of Higgs mass
and check, if your favourite NP model falls in these domains.

\bigskip
However, some caution is necessary:

\bigskip

1. if the mass of a new particle is
only slightly above $M_Z/2$
then the heavy mass expansion does not work;

\bigskip

2. the allowed domain of $S,U$ and $T$ depends on $M_H$.

\bigskip
To decouple or to nondecouple?

This is the first question you must ask analyzing NP.
The most famous example of NP with decoupling is SUSY.

Why do sleptons and squarks decouple? Because mass splitting
within SU(2) doublet is small, while from scalar fields
you can organize only vector current, which is conserved.
What about charginos and higgsinos? They form vector
multiplets (not chiral) which also decouples.

\bigskip
As a result the direct searches of the superpartners push lower
limits on their masses so high (hundreds of GeV) that their
contributions to rad.corr. are MOSTLY negligible.

Is there any relation between low energy SUSY and rad. corr.
except SUSY GUT?

Yes: in all the variants of SUSY the lightest Higgs boson mass
appeared to be less than 200 GeV, usually close to 100 GeV, which
nicely coincides with the values of $M_H$  obtained from rad.
corr.

\bigskip
4 generation \cite{MM}.

The simplest example of nondecoupled New Physics. It nondecouples just as the third generation with heavy top.
Mass of the neutral lepton $N$ should be
larger than $M_Z/2$ since $Z$ boson width allows only 3
light neutrino flavors.

Many new parameters: masses of
new particles and their mixing with three light generations.
For simplicity let us suppose that mixing is small.

\bigskip
At the next two slides the results of data fit
by the LEPTOP code performed by Alexander Rozanov in
summer 2008 are presented.

\bigskip
4 generation with 120 GeV Higgs

\begin{figure}[!htb]
\centering
\epsfig{file=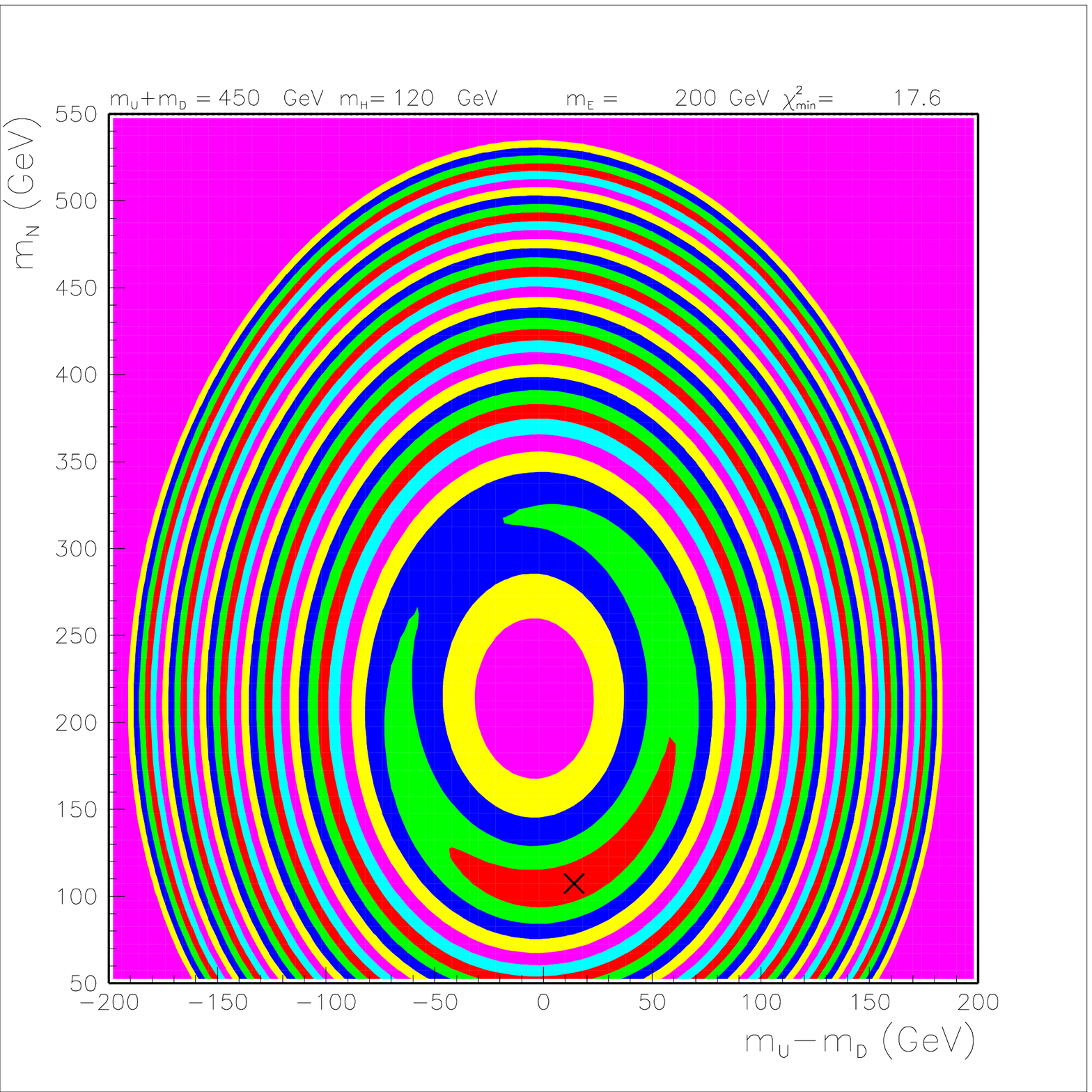,width=9cm}
\end{figure}

$m_E = 200$GeV, $m_U + m_D = 450$ GeV,
$\chi^2/d.o.f. = 17.6/11$, the quality of fit is
the same as in SM.

\bigskip
4 generation with 600 GeV Higgs
\begin{figure}[!htb]
\centering
\epsfig{file=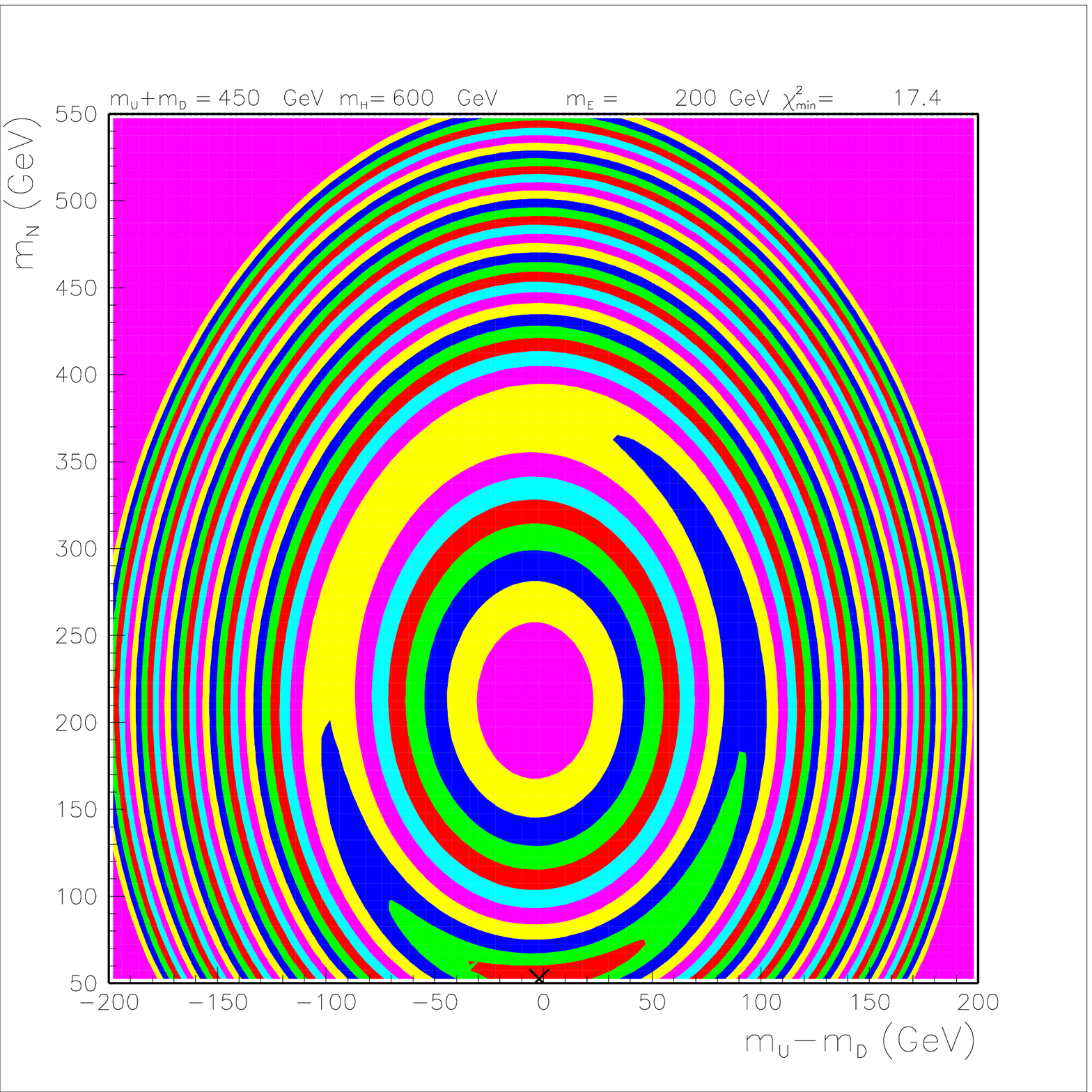,width=9cm}
\end{figure}

$m_E = 200$GeV, $m_U + m_D = 450$ GeV,
$\chi^2/d.o.f. = 17.4/11$, the quality of the fit is
the same as in SM.

\bigskip
So: Higgs is light ONLY in SM or if NP decouples.

\bigskip
Soon after LHC will start to produce physics the last
pages of Electroweak Interactions
will be written.

\bigskip
I am grateful to Victor Novikov, Lev Okun and Alexandre Rozanov
for many years of fruitful collaboration and to
Ahmed Ali and Mikhail Ivanov for the invitation 
to deliver lectures at School and for hospitality in Dubna.

This work was supported by Rosatom and grants RFBR 07-02-00021, 
RFBR 08-02-00494 and NSh-4568.2008.2.

\end{document}